\documentclass[12pt]{article}
\usepackage{pic03}
\usepackage{hyperref}
\usepackage{url}
\usepackage{graphicx}

\begin{document}

\title{\bf Nuclear Processes at Solar Energy}
\author{
Carlo Broggini (LUNA Collaboration)       \\
{\em Istituto Nazionale di Fisica Nucleare, via Marzolo 8, I-35131 Padova}}
\maketitle
% photograph of author
%  This is where we will insert a photograph. To see what it would look like,
%  uncomment the following lines.
%
%\begin{figure}[h]
%\begin{center}
%
% include photograph for proceeding version
%
%\includegraphics
%[height=4.5cm]{einstein.eps}
%
% insert a fixed vertical spacing instead for the ArXiv preprint
%
\vspace{4.5cm}
%
%\end{center}
%\end{figure}

\baselineskip=14.5pt
\begin{abstract}
LUNA, Laboratory for Underground Nuclear 
Astrophysics at Gran Sasso, is measuring fusion cross sections down to the energy 
of the nucleosynthesis inside stars. Outstanding results obtained up to now 
are the 
cross-section measurements
within the Gamow peak of the Sun of $^{3}He(^{3}He,2p)^{4}He$ and 
the $D(p,\gamma)^{3}He$. 
The former plays a big role 
in the proton-proton chain, largely affecting the calculated 
solar neutrino luminosity, whereas the latter is the reaction
that rules the proto-star life during the pre-main sequence phase.
The implications of such 
measurements will be discussed. 
Preliminary results obtained  last year 
on the study of $^{14}N(p,\gamma)^{15}O$,
the slowest reaction 
of the CNO cycle, will also be shown.
\end{abstract}
\newpage
\baselineskip=17pt

\section{Introduction}

Nuclear reactions that generate energy and synthesize elements take 
place inside the stars in a relatively narrow energy window:
the Gamow peak. In this region, which is in most
cases below 100 $keV$, far below the
Coulomb energy, the reaction cross-section 
$\sigma(\mathrm{E})$
drops almost exponentially 
with decreasing energy $E$ \cite{rolfs}:
\begin{equation}
\sigma(E)= \frac{S(E)}{E}\,{exp(-2\,\pi\, \eta)}, \label{yielddef1}
\end{equation}
where $S(E)$ is the astrophysical factor and
$\eta$ is the Sommerfeld parameter, given by
$2\, \pi\, \eta=31.29\, Z_1\, Z_2(\mu/E)^{1/2}$.
$Z_1$ and $Z_2$  are  the 
nuclear charges of the interacting particles  in  the  entrance channel,
$\mu$ is the reduced mass (in units of amu),  and  $E$  is  the  center 
of mass
energy  (in   units  of  keV). 
 
The extremely low value of the cross-section, ranging
from pico to femto-barn and even below, has always
prevented its measurement in a laboratory on the Earth's surface,
where the signal to background ratio would be too small because
of cosmic ray interactions. Instead, the observed energy dependence of the 
cross-section at high energies is extrapolated to the low energy
region, leading to substantial uncertainties.
In particular, there might be a change of the reaction mechanism 
or of the centrifugal barrier, or there might be the
contribution of narrow or sub-threshold resonances, all of which cannot 
be accounted by the extrapolation, but could completely 
dominate the reaction rate at the Gamow peak.

In addition, another effect can be studied at low energies: the electron
screening. The electron cloud surrounding the interacting nuclei 
acts as a screening potential, thus reducing the height of the 
Coulomb barrier and leading to a higher cross-section. The screening
effect has to be measured and taken into 
account in order to derive the bare 
nuclei cross-section, which is the input data to the models of stellar
nucleosynthesis.

In order to explore this 
new domain of nuclear astrophysics we have
installed two electrostatic 
accelerators underground in LNGS:
a 50 $kV$ accelerator \cite{nim1} and a 400 $kV$ one \cite{nim2}. 
The qualifying features of both
the accelerators are a very small beam energy spread and a very
high beam current even at low energy.
The accelerators are located 
in
two dedicated small rooms
of the Laboratori Nazionali del Gran Sasso
(LNGS), separated from other experiments by about 60 $m$ of 
rock. The mountain provides a natural shielding equivalent to
at least 3800 meters of water which reduces the muon and neutron fluxes 
by a 
factor $10^{6}$ and $10^{3}$, respectively. The $\gamma$ ray flux is like
the surface one,
but a detector can be more effectively shielded underground 
due to the suppression of the cosmic ray induced background.

\section{The $^{3}He(^{3}He,2p)^{4}He$ reaction}
The initial activity of LUNA has been focused 
on the $^{3}He(^{3}He,2p)^{4}He$ 
cross section measurement within the solar 
Gamow peak (15-27 $keV$).
Such reaction is a key one of the proton-proton chain.
A resonance at the thermal energy of the Sun 
was suggested 
long time ago  
\cite{fow72} \cite{fet72} to explain the observed $^{8}B$ 
solar neutrino flux:
it would decrease the relative contribution of the alternative 
reaction  
$^{3}He(\alpha,\gamma)^{7}Be$, which generates the branch responsible for
$^{7}Be$ and $^{8}B$
neutrino production in the Sun.
A narrow resonance with a peak S-factor 10-100 times the value extrapolated 
from high energy measurements could not be ruled out with the pre-LUNA data (such an
enhancement would be required to reduce the $^{7}Be$ and
$^{8}B$ solar neutrinos by a factor 2-3).
As a matter of fact, 
$^{3}He(^{3}He,2p)^{4}He$ 
cross section measurements stopped at the center of mass 
energy of 24.5 $keV$ 
($\sigma$=7$\pm$2 $pb$)\cite{kra87}, just at the
upper edge of the thermal energy region of the 
Sun. 

Briefly, the LUNA 
50 kV accelerator facility 
consisted of a duoplasmatron ion source,
an extraction/acceleration system, a double-focusing 90$^{\rm o}$ analyzing
magnet, a windowless gas-target system and a beam
calorimeter.
The beam energy spread was very small
(the source spread was less than 20 $eV$,  
acceleration voltage known with an accuracy of better than $10^{-4}$),
and
and the beam current was high even at low energy (about 300 $\mu$A
measurable with a 3$\%$ accuracy).
Eight thick (1 $mm$) silicon detectors of 
5x5 $cm^{2}$ area were
placed around the beam inside the 
target chamber, where there was 
a constant $^{3}He$ gas pressure of
0.5 $mbar$
(measured 
to an accuracy of better than 1\%).

The simultaneous detection of 2 protons has been 
the 
signature which unambiguously identified a
$^{3}He(^{3}He,2p)^{4}He$ fusion reaction (detection efficiency: 5.3$\pm$ 0.2$\%$,
Q-value of the reaction: 12.86 $MeV$).
No event fulfilling our selection criteria was detected
during a 23 day background run with a $^{4}He$ beam on
a $^{4}He$ target (0.5 $mbar$). 
\begin{figure}[htb]
\includegraphics[width=15cm]{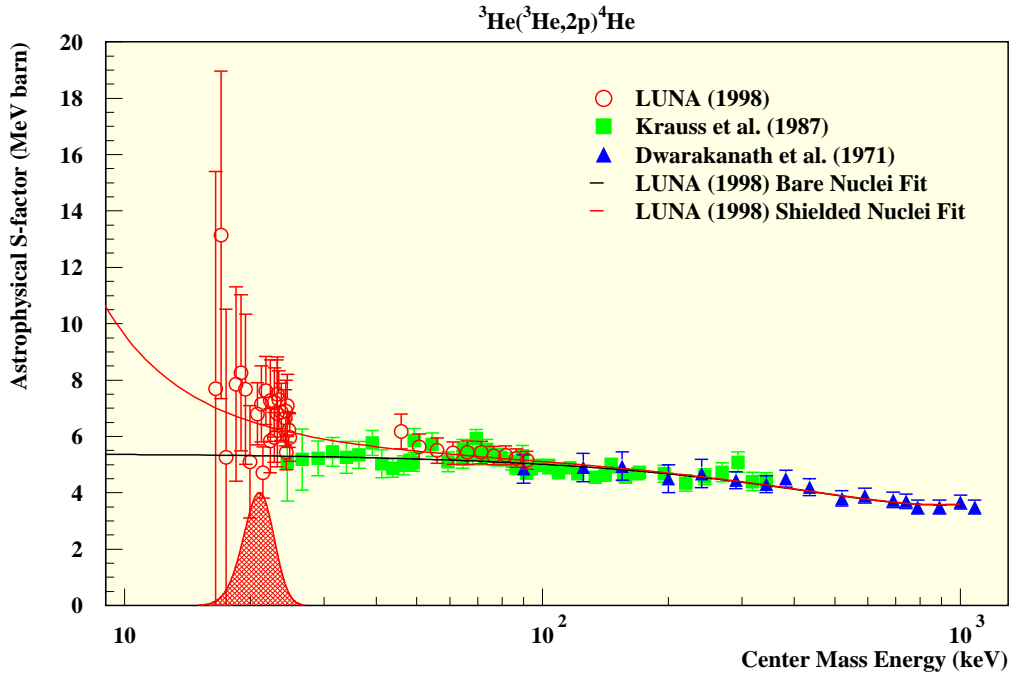}
\caption{The $^{3}He(^{3}He,2p)^{4}He$ astrophysical factor $S(E)$.
The position of the solar Gamow peak is also shown schematically.}
\label{fig1}
\end{figure}
Figure \ref{fig1} 
shows our results together with 
two existing measurements 
\cite{kra87}\cite{dwa71} 
of the 
astrophysical factor $S(E)$.
We point out that 
the cross section varies by more than two orders of magnitude in the 
measured energy range. At the lowest energy of 16.5 $keV$ it has 
the value of
0.02$\pm$0.02 $pb$,
which corresponds to
a rate 
of about 1 event/month, rather low even for the "silent" 
experiments of underground physics.

The LUNA result \cite{luna1} has shown 
that 
the $^{3}He(^{3}He,2p)^{4}He$ cross section
increases at the thermal energy of the Sun due to 
the electron 
screening effect but does not have any
narrow resonance.
Consequently, the astrophysical solution of the 
$^{8}B$ and $^{7}Be$ solar neutrino problem based on
its existence has been ruled out.

With $^{3}He(^{3}He,2p)^{4}He$ LUNA has provided the first 
cross section measurement
of a key reaction of the proton-proton chain
at the thermal energy of the Sun. In this way it has
also shown that, by going underground and by using 
the
typical techniques of low background physics, it is possible to 
measure 
nuclear cross sections down to the energy of the nucleosynthesis
inside stars.

\section{The D(p,$\gamma$)$^{3}$He reaction}
Inside the Sun, the D(p,$\gamma$)$^{3}$He reaction only effects the 
equilibrium abundance of deuterium. As a matter of fact, its cross section is
much higher than the one of the deuterium producer reaction  
p(p,e$^{+}\nu$)d. 

On the other hand, D(p,$\gamma$)$^{3}$He 
is the reaction which rules the life of the proto-stars before
they enter the main sequence phase. Reliable proto-star models predict 
that a star forms by accretion of interstellar material onto a
small contracting core. Until the temperature remains below 10$^{6}$
$K$, the main source of energy is the gravitational contraction.
When the temperature approaches 10$^{6}$ $K$ 
\begin{figure}[htb]
\includegraphics[width=15cm]{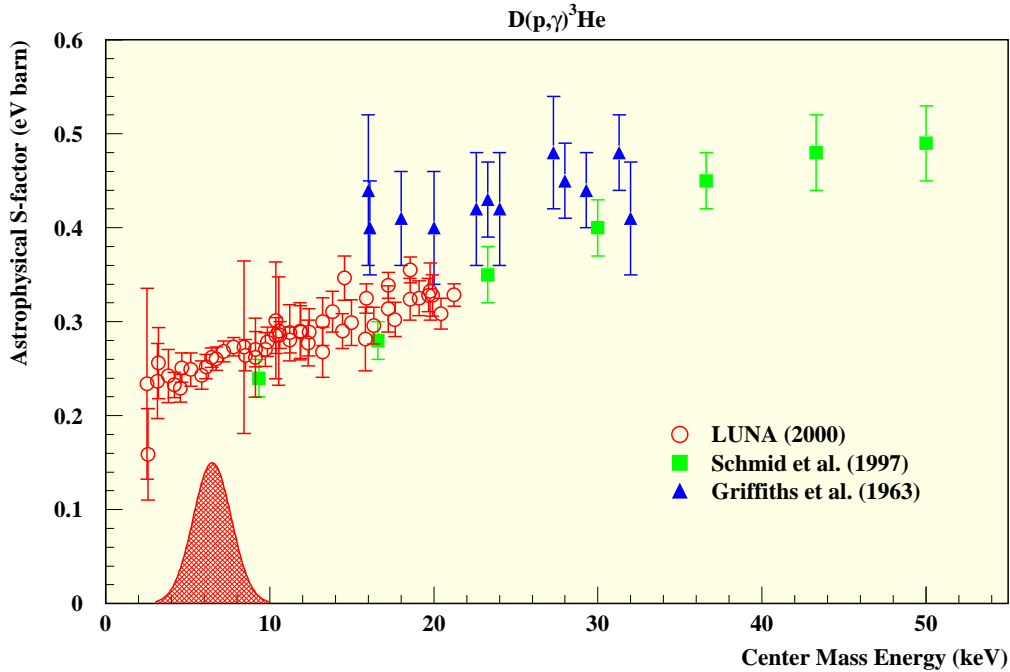}
\caption{The  D(p,$\gamma$)$^{3}$He astrophysical factor $S(E)$.
The position of the solar Gamow peak is also shown schematically.}
\label{fig2}
\end{figure}
the first "nuclear fire"
is switched on inside the star: 
the primordial deuterium is 
converted into $^{3}He$
via d(p,$\gamma$)$^{3}$He, thus providing 5.5 $MeV$ for each reaction.
The total amount of nuclear energy generated by this d-burning is comparable
with the whole gravitational binding energy of the star. The on-set
of d-burning slows down the contraction, increases the lifetime
of the star and freezes its observational properties until
the original deuterium is fully consumed. A reliable 
knowledge of the rate of d(p,$\gamma$)$^{3}$He down to a few $keV$
(the Gamow peak in a proto-star) is
a fundamental prerequisite for the proto-stellar models.

Finally, d(p,$\gamma$)$^{3}$He is also a cornerstone 
in the big-bang nucleosynthesis. Because of the deuterium 
"bottleneck" \cite{wein}, i.e. the photo-disintegration of deuterium,
the formation of $^{3}He$ is delayed until the temperature 
drops to about 8$\cdot$10$^{8} K$. As a consequence, 
the knowledge of the d(p,$\gamma$)$^{3}$He cross section at low energies
is necessary for the big-bang nucleosynthesis models.

The D(p,$\gamma$)$^{3}$He cross section measurement was made in LUNA
by using the 50 $kV$ accelerator connected to a differentially
pumped gas-target system designed to fit the characteristics of a large
BGO gamma ray detector \cite{nim3}. The BGO, a 28 $cm$ long cylinder
placed around the deuterium target, was detecting the 5.5 $MeV$ gamma with
a 70$\%$ efficiency.

Figure \ref{fig2} 
shows the LUNA results together with 
the only two existing measurements 
\cite{gri}\cite{sch} 
of the 
astrophysical factor $S(E)$ at low energy.
The cross section varies by more than three orders of magnitude in the 
measured energy range. At the lowest energy of 2.5 $keV$ it has 
the value of
9.2$\pm$4 $pb$,
which corresponds to
a rate 
of 50 events/day.

In figure \ref{fig2} we see one of the problems of nuclear astrophysics:
not only there are no measurements in the interesting energy region, but also the
extrapolations of the existing ones can have a significant
disagreement.

\section{The $^{14}N(p,\gamma)^{15}O$ reaction}
$^{14}N(p,\gamma)^{15}O$ is  the slowest reaction 
of the CNO cycle, the key one to know
the CNO solar neutrino flux, as well as to determine 
the age of the globular clusters,
the oldest components of the Milky Way.
As a matter of fact, the CNO solar neutrino flux 
depends almost linearly on this cross section
\cite{bahc}. The position of the 
Turn Off point in the Hertzsprung-Russel diagram of a globular cluster
is also determined by the value of the 
$^{14}N(p,\gamma)^{15}O$ cross section and it
gives the age of the cluster. As a matter of fact, a star
at the Turn-Off point is burning hydrogen in the shell through the CNO
cycle, it expands and, as a consequence, it has a decrease of the surface 
temperature.

The energy region studied 
so far in nuclear physics laboratories is well above the region
of interest for the CNO burning in astrophysical conditions
(20-80 $keV$). At solar energies the cross section of
$^{14}N(p,\gamma)^{15}O$
is dominated
by a sub-threshold resonance at -504~$keV$,
whereas at energies higher than 100~$keV$ it
is dominated by the 278 $keV$ resonance,
with
transitions to the ground-state of $^{15}O$ or to the 
excited states at energies of $5.18~MeV$, 
$6.18~MeV$ and $6.79~MeV$. 
According to Schr\"{o}der et al.
\cite{sch87}, who measured down to 0.2 $MeV$, the main
contribution to the total S-factor at zero energy,
S$(0)$,comes from the
transitions to the ground state of $^{15}O$ and to its
excited state at E$_{x}=6.79$~MeV. In particular, they
give
S$(0)=3.20\pm0.54$~$keV \cdot b$. On the other hand, Angulo et al. \cite{ang01} 
re-analyzed
Schr\"{o}der's experimental data using a R-matrix model
and they obtained
S$(0)=1.77\pm0.20$~$keV \cdot b$, which is a factor 1.7 lower than the
values used in the recent compilations. The difference
mainly comes from the different contribution of the
direct capture to the $^{15}$O 
ground state: Angulo et al.
have a value lower by a factor 19 than the one of Schr\"{o}der et al..
We underline that at the lowest energies  Schr\"{o}der
et al. give 
only upper limits to the cross section,
due to the presence of a strong 
cosmic ray background in the spectrum. 

In summary, new measurements of the $^{14}N(p,\gamma)^{15}O$
cross section at
energies $\mathrm{E}\leq200$~keV are 
strongly demanded. In particular it is necessary
to well measure the contribution  
of the direct capture to the ground state of $^{15}$O.
The peculiarities of the 400 $~kV$ 
LUNA facility \cite{nim2} are 
particularly well suited
for this study, where $\gamma$-rays with energy up to $\simeq~7.5~MeV$
have to be detected at very low count-rate
(Q-value of the reaction: 7.3 $MeV$). As a matter of fact, 
in such a measurement the
cosmic ray background has to be
strongly suppressed and ultra-low background
detectors have to be employed. In addition,
high beam intensities and
detectors with excellent energy
resolution have to be coupled to targets of high stability 
and purity, in order to minimize the beam-induced background. 

Due to the strong energy dependence of
the cross section, 
we carefully determined the uncertainties of
our accelerator:
$\pm 300$~$eV$ 
on the absolute energy from ${\rm E_p}=130$ to 400~$keV$,
proton energy spread of better than 100~$eV$ and long
term energy stability of 5~$eV$ per hour \cite{nim2}.

\begin{figure}[htb]
\includegraphics[width=15cm]{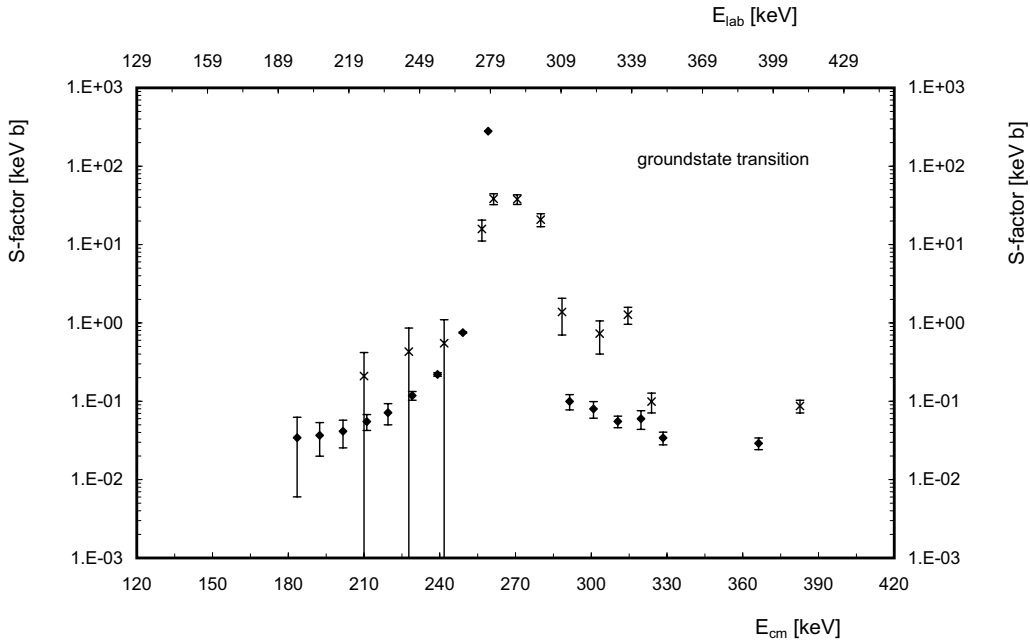}
\caption{Preliminary LUNA results for the ground-state transition
 of $^{14}N(p,\gamma)^{15}O$ (filled rhombus) and Schr\"{o}der results 
 (crosses).}
\label{fig3}
\end{figure}

The preliminary results shown in  figure \ref{fig3} have been obtained
with the beam passing through a Ta collimator 
and focused to a spot of about 1.5 $cm$ diameter on a $TiN$ target on $Ta$
backing.
A 126$\%$ HpGe detected the $\gamma$ rays 
from the reaction: it was placed at 55$^{o}$ from the
beam direction 
at about 1.5 $cm$ distance from the target. A detector with excellent
energy resolution is necessary in order to unambiguously separate the
different contribution to the cross section.

Figure \ref{fig3} is given just to show the data quality improvement
achieved thanks to the cosmic ray suppression. With only a small fraction 
of the data 
we already have an acceptable error in a region where the 
previous experiment could give only upper limits.
We are now completing the analysis of all the data we collected
in 2002 to cover the energy region down to about 100 $keV$.
With them it will be possible to separately know the
contribution to the cross section of both the direct capture to
the ground state of $^{15}O$ and to its excited state at 6.79 $MeV$.

\section{Present status and future directions}
In order to study the $^{14}N(p,\gamma)^{15}O$ reaction
down to the lowest energies
it is essential to have 
both a $\gamma$ ray detector with very high efficiency,
to compensate for the rapidly decreasing cross section,
and a very thin and pure $^{14}N$ target, to minimize 
the straggling on the energy loss and the beam induced background. 
This can be achieved with
the same 4$\pi$ BGO summing detector  
used in the measurement of
$D(p,\gamma)^{3}He$ and with a new
windowless gas target (a gas target is generally much more pure
than a solid one). Such a set-up has been constructed in 2001,
connected to the 400 $kV$ accelerator and fully tested before
summer this year. We have now started collecting the data.

There are two reactions already scheduled to be measured in the 
future:
$^{3}He(\alpha,\gamma)^{7}Be$ and
$^{25}Mg(p,\gamma)^{26}Al$.
$^{3}He(\alpha,\gamma)^{7}Be$ (Q-value: 1.6 $MeV$)
is the key reaction for the production
of $^{7}Be$ and $^{8}B$ neutrinos in the Sun. The joint effort of
all experiments on solar neutrinos and solar physics has finally 
cast light on the long-standing solar neutrino puzzle.
As a consequence, we can now go back to the original motivation of 
solar neutrino detection: the study of the Sun. The error on $S_{3,4}$,
about 16$\%$, is, at the moment, the main limitation to the extraction of
physics from the $^{8}B$ neutrino flux measurement. For instance, 
a 3-5$\%$
determination of $S_{3,4}$ would allow a study of the central region of the 
Sun with an accuracy better than the one given by helioseismology.

$^{25}Mg(p,\gamma)^{26}Al$ (Q-value: 6.3 $MeV$) is 
a  reaction of the $MgAl$ cycle. Its low energy measurement would be
important for two reasons: the study of the nucleosynthesis of
the elements with mass number between 24 and 27 and the 
$\gamma$ ray astronomy. As a matter of fact, $^{26}Al$ is a
radioactive nucleus, with 7.4$\cdot 10^{5}$ year half-life.
Its decay gives rise to a 1.8 $MeV$ gamma ray: one of the 
key line of $\gamma$ astronomy. There is now a full sky map
at 1.8 $MeV$, provided by the NASA Compton Gamma-Ray Observatory.
The ESA INTEGRAL Observatory is going to improve the picture of
the 1.8 $MeV$ sky in
the near future. The study of the reaction producing $^{26}Al$ 
is essential to understand such a sky.

\section{Conclusions}
LUNA started its activity almost 10 years ago 
in order to explore the new domain of nuclear astrophysics 
at low energy. During these years it has proved that,
by going underground 
and by using the typical techniques of low background 
physics, it is possible to measure nuclear cross sections down 
to the energy of the nucleosynthesis inside stars.

In particular, we have provided the only existing 
measurements of important fusion reactions
within the Gamow peak of the Sun:
$^{3}He(^{3}He,2p)^{4}He$  
and $D(p,\gamma)^{3}He$. The results on 
$^{3}He(^{3}He,2p)^{4}He$ have shown that nuclear
physics was not the origin of the solar neutrino puzzle.

We are now measuring 
$^{14}N(p,\gamma)^{15}O$, the key reaction 
of the CNO cycle. After this we will study 
$^{3}He(\alpha,\gamma)^{7}Be$ and
$^{25}Mg(p,\gamma)^{26}Al$. The former 
is the key reaction for the production
of $^{7}Be$ and $^{8}B$ neutrinos in the Sun,
whereas the latter is essential to understand the $\gamma$ 
sky at 1.8 $MeV$.

\end{document}